# Enhancement of Superconductivity Linked with Linear-in-Temperature/Field Resistivity in Ion-Gated FeSe Films

Xingyu Jiang[1,2#], Mingyang Qin[1,2#], Xinjian Wei[1#], Zhongpei Feng[1,5], Jiezun Ke[3,4], Haipeng Zhu[3,4], Fucong Chen[1,2], Liping Zhang[1], Li Xu[1,2], Xu Zhang[1,2], Ruozhou Zhang[1,2], Zhongxu Wei[1,2], Peiyu Xiong[1], Qimei Liang[6], Chuanying Xi[6], Zhaosheng Wang[6], Jie Yuan[1,2,5], Beiyi Zhu[1], Kun Jiang[1,2], Ming Yang[3,4], Junfeng Wang[3,4], Jiangping Hu[1,2], Tao Xiang[1,2], Brigitte Leridon[7], Rong Yu[8], Qihong Chen[1,2,5*], Kui Jin[1,2,5*] and Zhongxian Zhao[1,2,5]

[1]Beijing National Laboratory for Condensed Matter Physics, Institute of Physics, Chinese Academy of Sciences, Beijing 100190, China

[2]School of Physical Sciences, University of Chinese Academy of Sciences, Beijing 100049, China

[3]Wuhan National High Magnetic Field Center, Huazhong University of Science and Technology, Wuhan 430074, China

[4]School of Physics, Huazhong University of Science and Technology, Wuhan 430074, China

[5]Songshan Lake Materials Laboratory, Dongguan, Guangdong 523808, China

[6]Anhui Province Key Laboratory of Condensed Matter Physics at Extreme Conditions, High Magnetic Field Laboratory of the Chinese Academy of Sciences, Hefei 230031, Anhui, China

[7]LPEM, ESPCI Paris, PSL Research University, CNRS, Sorbonne Université, 75005 Paris, France

[8]Department of Physics and Beijing Key Laboratory of Opto-electronic Functional Materials and Micro-nano Devices, Renmin University of China, Beijing 100872, China

#These authors contributed equally to this work.
*email: qihongchen@iphy.ac.cn; kuijin@iphy.ac.cn

**Abstract**

Iron selenide (FeSe) – the structurally simplest iron-based superconductor, has attracted tremendous interest in the past years. While the transition temperature ($T_c$) of bulk FeSe is ~ 8 K, it can be significantly enhanced to 40 – 50 K by various ways of electron doping. However, the underlying physics for such great enhancement of $T_c$ and so the Cooper pairing mechanism still remain puzzles. Here, we report a systematic study of the superconducting- and normal-state properties of FeSe films via ionic liquid gating. With fine tuning, $T_c$ evolves continuously from below 10 K to above 40 K; *in situ* two-coil mutual inductance measurements unambiguously confirm the gating is a uniform bulk effect. Close to $T_c$, the normal-state resistivity shows a linear dependence on temperature and the linearity extends to lower temperatures with the superconductivity suppressed by high magnetic fields. At high fields, the normal-state magnetoresistance exhibits a linear-in-field dependence and obeys a simple scaling relation between applied field and temperature. Consistent behaviors are observed for different-$T_c$ states throughout the gating process, suggesting the pairing mechanism very likely remains the same from low- to high-$T_c$ state. Importantly, the coefficient of the linear-in-temperature resistivity is positively correlated with $T_c$, similarly to the observations in cuprates, Bechgaard salts and iron pnictide superconductors. Our study points to a short-range antiferromagnetic exchange interaction mediated pairing mechanism in FeSe.

**Introduction**

The $T_c$ (transition temperature) of high-temperature superconductors can be modulated by tuning various physical parameters such as carrier concentration. Determining the key ingredient that drives the $T_c$ evolution is of great importance for understanding the superconducting pairing mechanism. Among the high-temperature superconductors, iron selenide (FeSe) provides an ideal platform for investigating the physics behind $T_c$ evolution owing to its simple structure and widely tunable superconducting properties [1]. While bulk FeSe shows a $T_c$ of ~ 8 K [2], by different ways of electron doping e.g. surface K dosing [3–5] and alkali-metal/molecule interlayer intercalation [6–10], $T_c$ can be enhanced to 40 – 50 K. FeSe-based superconductors with high $T_c$'s share a common feature that the Fermi surface (FS) topology exhibits electron pockets only [3,4,11,12], while the FS of bulk FeSe is composed of both hole (**Γ** point, Brillouin zone center) and electron pockets (**M** point, Brillouin zone corner) [13,14]. Seemingly the change of FS topology, i.e. the Lifshitz transition, plays an important role in the $T_c$ enhancement [15].

To determine the microscopic mechanism that depicts the superconductivity of FeSe, it is of first priority to figure out how the electronic properties evolve during $T_c$ enhancement. For this purpose, ionic liquid gating serves as a promising tool. Using liquid as the gating media to replace the traditional oxides, this method is capable of modifying the doping level of a material to a large extent and accordingly changing its electronic properties [16–21]. More importantly, it can be integrated with various *in situ* characterization techniques, e.g. electrical transport, magnetic response and thermoelectric measurements, therefore can provide valuable information about the evolution of the superconducting- and normal-state properties. Via this method, previous studies succeeded in inducing high-$T_c$ (~ 40 K) superconductivity in FeSe thin films [22–27]. In this work, we report a continuous tuning of $T_c$ from below 10 K to above 40 K in FeSe films by ionic liquid gating. A two-coil mutual inductance technique is integrated, which enables the characterization of the superconducting volume. The result unambiguously confirms that the gating is a uniform bulk effect, ensuring a reliable analysis of the electronic properties. For superconducting states with different $T_c$'s during ionic gating, a linear-in-$T$ and linear-in-$H$ resistivity is observed at high magnetic fields and low temperatures, which exhibits a simple scaling relation with field and temperature. The coefficient of the linear-in-$T$ resistivity is positively correlated with $T_c$, similarly to the observations in cuprates [28] and Bechgaard salts [29,30], etc. Our results exemplify the important role of quantum fluctuations in governing the Cooper pairing of high-temperature superconductors.

**Methods**

The FeSe films, with thicknesses of a few hundred nanometers, were grown on $CaF_2$, $SrTiO_3$ or LiF substrates by pulsed-laser deposition [31]. For ionic gating, the FeSe film and Pt gate electrode were covered by the ionic liquid: *N*,*N*-diethyl-*N*-(2-methoxyethyl)-*N*-methylammonium bis (trifluoromethylsulphonyl) imide (DEME-TFSI). The temperature dependence of resistance (*R-T*) was measured using a d.c. four-terminal method, in either a Quantum Design Physical Property Measurement System (PPMS) or Montana cryostation. The sample was patterned into standard Hall-bar geometries for the Hall resistivity measurements. The magnetic response during gating was detected by *in situ* two-coil mutual inductance technique [32]. High-field transport (*ab*-plane magnetoresistance) during gating was measured either in steady High Magnetic Field Facilities (up to ~ 33 T) at High Magnetic Field Laboratory, Chinese Academy of Sciences, or in a pulsed magnet up to 50 T with pulse duration of 150 ms at the Wuhan National High Magnetic Field Center (China).

**Experimental Results**

The device configuration of ionic gating on a FeSe film is shown in Fig. 1(a). By tuning the gate voltage ($V_G$), gating temperature ($T_G$) and duration time, the superconducting state of FeSe films can be effectively modulated. Figure 1(b) summarizes the *R-T* curves at different stages of ionic gating. Hereafter, we use the low-$T_c$ and high-$T_c$ states to denote superconducting states with $T_c$ ~ 10 K and $T_c$ ~ 40 K, respectively. Overall, the superconductivity shows a continuous evolution from

low- to high-$T_c$ state. It should be noted that in Fig. 1(b) different $R$-$T$ curves are collected from different successive gating experiments. As a matter of fact, obtaining a superconducting state with an intermediate $T_c$ value is experimentally challenging, since it only exists in a narrow window of the gating process. For each experiment, only a few intermediate $T_c$ values were obtained. However, the repetitive observations of different $T_c$ values in different experiments confirm the presence of intermediate superconducting states. Figure 1(c) shows the statistics of the $T_c$ values extracted from 10 different gating experiments. Here, $T_c$ is defined as the onset temperature of the superconducting transition, as demonstrated in Fig. 1(b). As can be seen, $T_c$ almost covers all values from 10 to 50 K, suggesting that the evolution of superconductivity is a smooth and continuous process.

Before proceeding to analyze the evolution of electronic properties, one might naturally wonder whether the continuous evolution of $T_c$ represents a bulk property of FeSe. It is known that the homogeneity of ionic gating is a tricky issue. As a result, inhomogeneous gating could cause phase separation and an intermediate $T_c$ value between the low- and high-$T_c$ states. In the electrical transport measurement, once a percolating superconducting channel is formed with relatively high $T_c$, it will shunt the other low-$T_c$ or normal conducting channels. Therefore, the resistive transition offers information about the superconducting channel with the highest $T_c$, but fails to reveal the uniformity of the superconducting state.

To better understand the evolution of the superconducting state and its corresponding normal state, we developed an *in situ* two-coil mutual inductance (TCMI) technique combined with ionic gating to detect the magnetic response of the whole film [32], thus the superconducting volume of the high-$T_c$ state can be characterized. Figure 2(a) shows a schematic illustration of the TCMI device, in which an alternating current $I_{in}$ applied to the drive coil induces an alternating voltage $V_p$ in the pick-up coil. Details about the TCMI measurement are described in Section 3 of Supplemental Material. When the FeSe film enters the superconducting state at low temperatures, the Re$V_p$ (real part of the pick-up signal) undergoes a sudden drop corresponding to the diamagnetic response of the sample. From the Re$V_p$-$T$ curve, we can distinguish whether the gating is a bulk effect or there exists phase separation. For example, Figure 2(b) shows a typical measurement of the Re$V_p$-$T$ curves during ionic gating in the case of inhomogeneous gating. Initially, there is only one drop at ~ 9 K in both the Re$V_p$-$T$ (black) and $R$-$T$ [inset of Fig. 2(b)] curves, corresponding to a uniform superconducting state. After ionic gating ($V_G$ = 4.5 V, $T_G$ = 250 K), several drops are observed (the red curve), suggesting the presence of more than one superconducting phase due to inhomogeneous gating. This information is completely concealed in the electrical transport measurements, where only one resistive transition is observed for the high-$T_c$ state [the red curve in the inset of Fig. 2(b)].

To improve the uniformity, we raised the gating temperature to $T_G$ = 295 K. By gradually increasing the gate voltage, a uniform evolution from low- to high-$T_c$ state can be realized in a single gating sequence, as shown in Fig. 2(c). For each curve, there is only one main transition visible in Re$V_p$, implying a uniform bulk modulation from low- to high-$T_c$ phase. It should be noted that Re$V_p$ of the high-$T_c$ state at the lowest measured temperature is slightly higher than that of the low-$T_c$ state, which is possibly due to a moderate electrochemical etching effect at a higher gating temperature. Therefore, raising the gating temperature improves the uniformity of the superconducting state but also enhances the risk of electrochemical etching. However, the magnitude of the difference is so small that the etching effect, even if present, is not the key factor for the $T_c$ enhancement. More details about the mechanism of ionic liquid gating on FeSe can be found in Section 4 of Supplemental Material. Importantly, the parallel shift of the superconducting transition in Re$V_p$-$T$ curves to higher temperatures unambiguously confirms that the gating is homogeneous and the continuous evolution of $T_c$ reflects an intrinsic bulk property of FeSe.

Now that we succeed in controlling a continuous tuning of $T_c$, we turn to analyzing the corresponding normal-state transport properties. During the gating process, the Hall signal changes from positive at low-$T_c$ phase to negative at high-$T_c$ phase (Fig. S4 of Supplemental Material), indicating the change of the FS from the coexistence of electron and hole pockets to electron pockets only (i.e. the Lifshitz transition), consistently with previous reports [3,23]. Although it is difficult to determine exactly at which $T_c$ the Lifshitz transition occurs, the continuous evolution of $T_c$ suggests a smooth evolution of the superconducting state, i.e. no abrupt variation of $T_c$ with the change of FS topology. Therefore the FS nesting is most likely not the primary factor for electron pairing in FeSe.

Figure 3(a) displays the temperature dependence of resistivity ($\rho$-$T$) for one gating sequence. The magnetic response was monitored simultaneously with the electrical transport measurement (Fig. S3 of Supplemental Material), to guarantee that the gating is a uniform bulk modulation and the electrochemical etching effect is negligible. For each curve, there is a region above $T_c$ where the resistivity varies linearly with temperature. This linear-in-$T$ resistivity can extend to lower temperatures when superconductivity is suppressed by high magnetic fields, as can be seen in Figs. 3(b)-(d). For a pristine FeSe film ($T_c \sim 11$ K), with superconductivity suppressed by a steady magnetic field of 33 T, the resistance shows a perfect linear dependence on temperature spanning over a decade from 20 K to the lowest measured temperature 2 K [Fig. 3(b)]. For ion-gated FeSe films with higher $T_c$'s, pulsed magnetic fields up to 50 T were applied to suppress the superconductivity and explore the normal-state properties below $T_c$. For the states with $T_c \sim 35$ K [Fig. 3(c)] and $\sim 43$ K [Fig. 3(d)], even though the magnetic field of 50 T is not enough to completely suppress the superconductivity, the linear resistance extends to much lower temperatures. It is reasonable to infer that under a higher magnetic field, the linear-in-$T$ behavior will persist to the low-temperature limit. This conjecture is indeed corroborated in another FeSe-based superconductor – (Li, Fe)OHFeSe [33], which exhibits very similar superconducting properties to ion-gated FeSe. When the superconductivity is suppressed by a magnetic field of 70 T, the resistivity of (Li, Fe)OHFeSe shows a linear temperature dependence down to 1.4 K [Fig. S6 of Supplemental Material].

For materials showing linear-in-$T$ resistivity, a concomitant linear-in-field magnetoresistance (MR) is commonly observed, which follows a scaling relation with applied magnetic field and temperature [34–36]. Similar phenomenon is indeed observed in our FeSe films, as shown in Fig. 4(a). The resistivity is normalized to the value at 250 K [i.e. $\tilde{\rho} = \rho(T,H)/\rho(250\text{ K}, 0\text{ T})$] to remove the influence of geometric factors. At low temperatures, with superconductivity suppressed by high magnetic fields, the normal-state resistivity exhibits an $H$-linear dependence without any sign of saturation to the highest measured field. At elevated temperatures, a $H^2$-dependent resistivity develops at low fields, resembling what is expected in a conventional metal where the MR is proportional to $H^2$. Yet for an intermediate temperature the MR remains $H$-linear at high fields. With further increase of temperature, the $H^2$-term continues to grow and eventually occupies the whole measured field range. To capture the scaling behavior in the MR, we performed a scaling analysis similar to that proposed for an iron pnictide [34] and a cuprate [35]. Figure 4(b) shows $(\tilde{\rho} - \tilde{\rho}_0)/T$ versus $\mu_0 H/T$, where $\tilde{\rho}_0$ is the residual resistivity obtained by extrapolating the $H$-linear $\tilde{\rho}$ at finite temperatures to $H = 0$ and $T = 0$. As can be seen, the normal-state resistivity collapses onto a single curve. Highly consistent behaviors are also observed for ion-gated samples with high $T_c$'s, as shown in Figs. S7 and S8 of Supplemental Material. We attempted to fit the data employing functions $(\tilde{\rho} - \tilde{\rho}_0)/T \propto [1 + b(\mu_0 H/T)]$ (black dashed line, Ref. [35]) and $(\tilde{\rho} - \tilde{\rho}_0)/T \propto \sqrt{1 + c(\mu_0 H/T)^2}$ (red dashed line, Ref. [34]), respectively, where $b$ and $c$ are numeric parameters. Both forms fit the data well at high $H/T$ ratios, while significant deviation is observed for the latter at low $H/T$ ratios. This result is in good agreement with that reported in La$_{2-x}$Ce$_x$CuO$_4$ [35]. As a consequence, the resistivity can be written as $\rho - \rho_0 \propto Ak_B T + C\mu_B \mu_0 H \equiv \varepsilon(T, H)$, which means the resistivity is proportional to the linear sum of thermal and magnetic field energies at low temperatures. It strongly suggests that there is a scale-invariant region (i.e., lack of an intrinsic energy scale) and that the linear-in-$H$ resistivity behavior has the same origin as the linear-in-$T$ resistivity behavior.

Note that the linear-in-$T$ resistivity has also been observed in other FeSe-based superconductors, e.g. (TBA)$_x$FeSe [9], Li$_x$(C$_3$N$_2$H$_{10}$)$_{0.37}$FeSe [10] and FeSe$_{1-x}$S$_x$ [37], but has not been systematically discussed in this material family. In Fig. 5, we extract the information of the linear-in-$T$ resistivity during the $T_c$ evolution of FeSe by ionic gating. The linear resistivity coefficient $A$, obtained by fitting the linear region of $\rho$-$T$ curves with the form $\rho = \rho_0 + AT$, exhibits a clear positive correlation with $T_c$ [Fig. 5(a)]. We also extract the upper-bound temperature $T_1$, above which the dependence of resistivity deviates from the linear behavior and crosses over to a $T^2$-dependence (Fig. S9 of Supplemental Material). Below $T_1$, the linear-in-$T$ electron scattering mechanism dominates the electrical transport. In addition to the positive correlation between $A$ and $T_c$, a positive correlation between $T_1$ and $T_c$ is observed [Fig. 5(b)]. Higher $T_1$ corresponds to a larger energy scale, implying that the strengthening of the microscopic mechanism that leads to the linear-in-$T$ resistivity also results in the enhancement of superconductivity.

**Discussions and Summary**

Linear-in-$T$ resistivity, known to be the key feature of the strange-metal phase [38], is commonly observed in unconventional superconductors [28–30,38–42], yet its origin is under hot debate. In hole-doped cuprates, the notion of Planckian dissipation [43] – which sets a fundamental bound on the scattering rate $1/\tau$ in condensed matter systems – has been invoked to describe the linear-in-$T$ resistivity. However, this notion cannot explain the crossover of the resistivity in FeSe from $T$-linear to $T^2$-dependence at $\sim 60 - 90$ K (Fig. S9 of Supplemental Material), above which the scattering rate exceeds the Planckian limit, i.e., the resistivity is higher than the value extrapolated from the low-temperature linear dependence. A similar constraint has been found in electron-doped cuprates, where the crossover temperature is even lower ($\sim 25$ K) [38].

Alternatively, it has been proposed in some other systems that the linear-in-$T$ scattering could stem from two-dimensional (2D) or disordered three-dimensional (3D) antiferromagnetic (AF) spin fluctuations [44,45]. This scenario can explain the electron pairing and strange-metal phase in electron-doped cuprates [28] and Bechgaard salts (TMTSF)$_2$X (X = PF$_6$, ClO$_4$) [29,30]. These systems share a common feature with the ion-gated FeSe in this study, that is, a universal positive correlation between the $T$-linear coefficient $A$ and $T_c$ [Fig. 5(a)]. FeSe has been found, by inelastic neutron scattering, to have a large fluctuating magnetic moment ($<m^2> =$

5.19 $\mu_B^2$/Fe) [46], and its spin excitations can be understood by models of quasi-localized magnetic moments with short-range AF exchange interactions [47–49]. Accordingly, we propose that the linear-in-$T(H)$ resistivity and superconductivity in FeSe are driven by the short-range AF exchange interactions, similar to the cases in electron-doped cuprates and Bechgaard salts. In this scenario, the superconducting pairing amplitude is determined by the strength of electron correlations [47,50], while the FS nesting is not the primary factor. With ionic gating, the electron correlations are gradually enhanced, resulting in the enhancement of superconductivity. A direct consequence of this conclusion is that the low- and high-$T_c$ states can be understood within the same picture of electron pairing, in good agreement with the continuous evolution of $T_c$ and linear-in-$T(H)$ behaviors observed at all different-$T_c$ states throughout the gating process. The strong coupling scenario is highly consistent with a previous angle-resolved photoemission spectroscopy study on FeSe [4], where the electron correlations are evidently enhanced with electron doping, accompanying the increase of $T_c$. It is also supported by the observations of an approximately linear relation between $T_c$ and superfluid density, and a large and non-constant ratio of superconducting energy gap to $T_c$ with surface K dosing on FeSe [5].

In optimally hole-doped cuprates the linear-in-$T$ resistivity was actually identified very early on as a smoking gun for the breaking down of Landau quasiparticle concept and the corresponding fermions were termed marginal Fermi liquid [51]. It was shown recently that a new class of quantum-critical fluctuations that are orthogonal in space and time, determine a critical spectrum fully consistent with this marginal Fermi liquid behavior, and the scattering of fermions on these fluctuations gives rise to linear-in-$T$ resistivity [52]. The AF quantum fluctuations might actually under some circumstances map onto a quantum XY model [53] and yield the same kind of space-and-time-separated quantum fluctuations, resulting in the linear-in-$T$ resistivity. Namely, this linear-in-$T$ resistivity can be related to the same class of quantum fluctuations but with two different microscopic mechanisms for hole-doped cuprates and iron-based superconductors. It has been argued that the scaling of the resistivity as a function of applied magnetic field $H$ and temperature $T$ in $BaFe_2(As_{1-x}P_x)_2$ is an important signature of the putative quantum critical point (QCP) beneath the superconducting dome. Similar scale-invariant behavior between $H$ and $T$ observed in electron- and hole-doped cuprates also corroborate this idea. For iron pnitcides and electron-doped cuprates, the QCPs are most likely associated with long- or short-range AFM order, while for hole-doped cuprates, the QCP resides at the end of a pseudogap phase of unknown origin. However, in FeSe systems, there is no long-range AFM order. In the temperature-doping phase diagram [4], a nematic phase intersects with the superconducting dome at a doping level where $T_c$ boosts. In the temperature-pressure phase diagram [54], a spin density wave order develops above the superconducting dome. Whether these phenomena are correlated with the quantum critical behavior and the underlying microscopic mechanism require further investigation. Nonetheless, the experimental observations have demonstrated a vivid correlation between the linear-in-$T$ scattering and $T_c$, pointing to a universal phenomenon in iron-based superconductors, cuprates and Bechgaard salts.

In summary, we realized a continuous tuning of $T_c$ in FeSe films from below 10 K to above 40 K. Further combining the TCMI technique, we confirmed the gating is a bulk modulation of the superconducting properties, which allows for a reliable analysis of the normal-state properties for the corresponding superconducting phase. Linear-in-$T$ dependence of the normal-state resistivity is thus found at low temperatures, and remarkably, the slope is positively correlated with $T_c$, pointing to a universal phenomenon in unconventional superconductor systems like cuprates and Bechgaard salts. Moreover, the linear-in-$T$ resistivity behavior is accompanied by a linear-in-$H$ normal-state resistivity at high fields and low temperatures, and there is a scale-invariant region where the resistivity follows a simple scaling relation with field and temperature. Our work suggests the spin fluctuations associated with short-range AF exchange interactions presumably play an essential role in the scattering and electron pairing of FeSe throughout the whole tuning process.

**Acknowledgements**

We thank Hongyi Xie, Gang Xu, Zhenyu Zhang, Guangming Zhang, Kai Liu, Xianxin Wu, Ge He and Lei Liang for stimulating discussions. This work was supported by the National Key Basic Research Program of China (2017YFA0302902, 2016YFA0300301, 2017YFA0303003, and 2018YFB0704102), the National Natural Science Foundation of China (11674374, 11927808, 11888101, 11834016, 118115301, 119611410, 11961141008 and 11874359), the Strategic Priority Research Program (B) of Chinese Academy of Sciences (XDB25000000, XDB33000000), the Key Research Program of Frontier Sciences, CAS (QYZDB-SSW-SLH008 and QYZDY-SSW-SLH001), CAS Interdisciplinary Innovation Team, Beijing Natural Science Foundation (Z190008). R.Y. was supported by the National Science Foundation of China Grant number 11674392, the Fundamental Research Funds for the Central Universities and the Research Funds of Renmin University of China Grant number 18XNLG24, and the Ministry of Science and Technology of China, National Program on Key Research Project Grant number 2016YFA0300504.

**Author contributions**

Q.H.C., X.Y.J., M.Y.Q. and X.J.W. performed the electrical transport and two-coil mutual inductance measurements. M.Y.Q. and R.Z.Z. designed the two-coil mutual inductance measurement device. Q.H.C. and M.Y.Q. performed the high magnetic field measurement, with the help of M.Y., J.Z.K., H.P.Z., Q.M.L., C.Y.X., Z.S.W., L.X., R.Z.Z. and P.Y.X. FeSe films were grown and characterized by Z.P.F., F.C.C. and L.P.Z.. Q.H.C., X.Y.J. and K.J. wrote the manuscript. All the authors discussed and contributed to the manuscript. Q.H.C., K.J. and Z.X.Z. conceived the project.

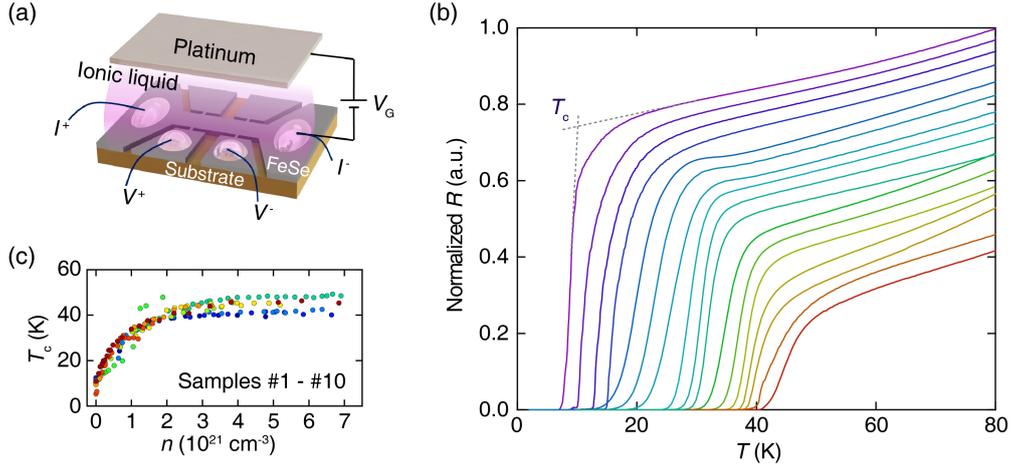

FIG. 1. Continuous evolution of superconductivity by ionic gating. (a) Schematic illustration of the ionic gating device. (b) Temperature dependence of the resistance ($R$-$T$) at different stages of ionic gating. Different curves are taken from different gating experiments. The resistance is normalized in a manner that the trend roughly reflects the relative change of the absolute values in one successive gating experiment. For a complete set of $R$-$T$ curve in one successive gating experiment, see Figs. S2 and S3 in Supplemental Material. (c) The $T_c$ as a function of electron doping concentration, extracted from 10 different gating experiments, which almost covers all values from 5 to 50 K. It should be noted that the electron doping level is calculated by the integral of leakage current during gating for samples #5, #7 and #10. Others are mapped to the plot with rough estimations, due to the irregular geometries of different samples. Therefore, the doping concentration is of low accuracy thus is not suitable for any quantitative analysis. However, the continuous evolution of $T_c$ is vividly demonstrated.

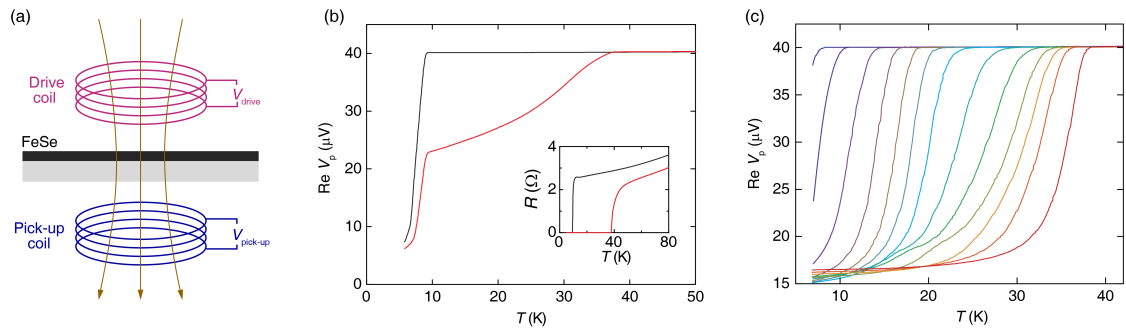

FIG. 2. *In situ* magnetic response measurement by the TCMI technique. (a) Schematic illustration of the TCMI device. (b) Typical temperature dependences of the real part of the pick-up signal $V_p$, before (black curve) and after (red curve) ionic gating. Inset: The *R-T* curves corresponding to the two superconducting states in the main panel. (c) The evolution of Re$V_p$-*T* curves in one successive gating sequence. From left to right, the gating voltages are: 0, 0.5, 1.5, 3.0, 3.2, 3.4, 3.6, 3.8, 4.0, 4.2, 4.4, 4.4 and 4.5 V, respectively. The gating temperature is 295 K, and for each gating voltage, the duration time is 3 hours.

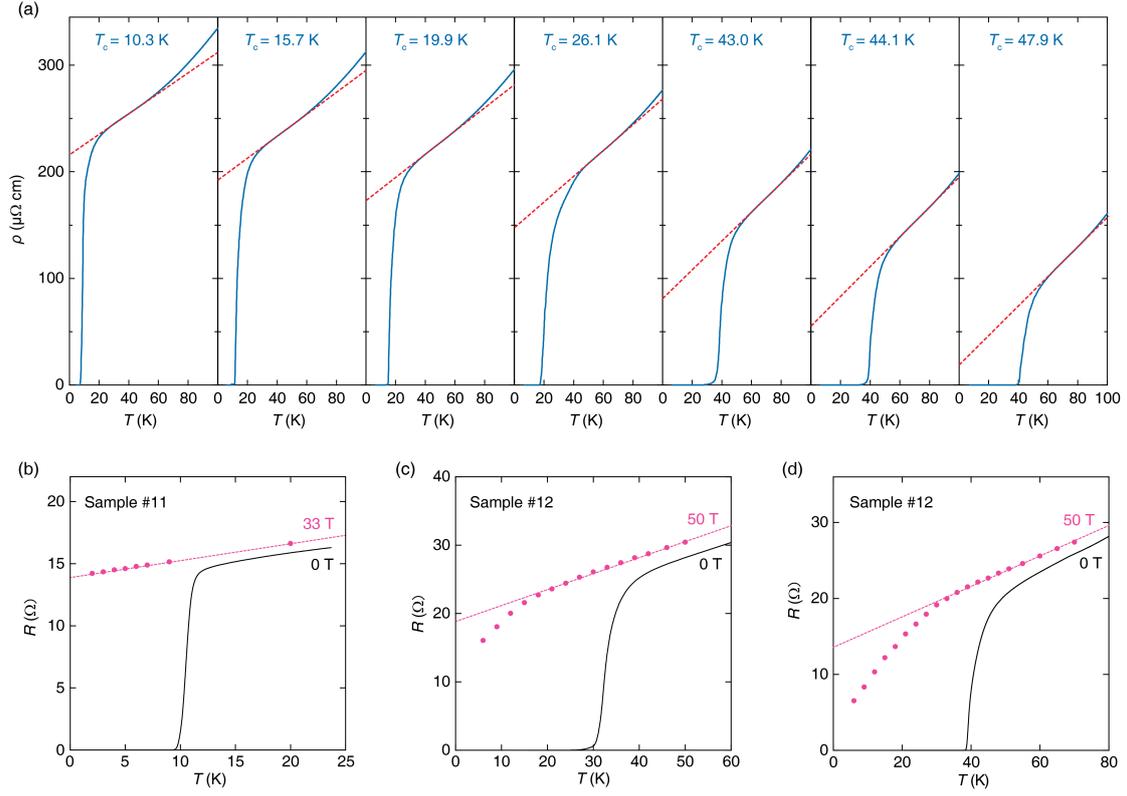

FIG. 3. The linear-in-$T$ resistivity. (a) Temperature dependence of the resistivity in one successive gating sequence for sample #5, displaying linearity in the vicinity of $T_c$. The solid lines are experimental data, and the red dash lines are the fitting results to the form $\rho = \rho_0 + AT$. (b)-(d) Zero-field (black lines) and high-field (pink dots) $R$-$T$ of (b) a pristine FeSe film (sample #11) with $T_c \sim 11$ K, a gated film (sample #12) with $T_c$ (c) $\sim 35$ K and (d) $\sim 43$ K, respectively. The pink dashed lines are guidelines for the linear-in-$T$ resistance. The magnetic field is applied perpendicular to the FeSe film, i.e., $B//c$.

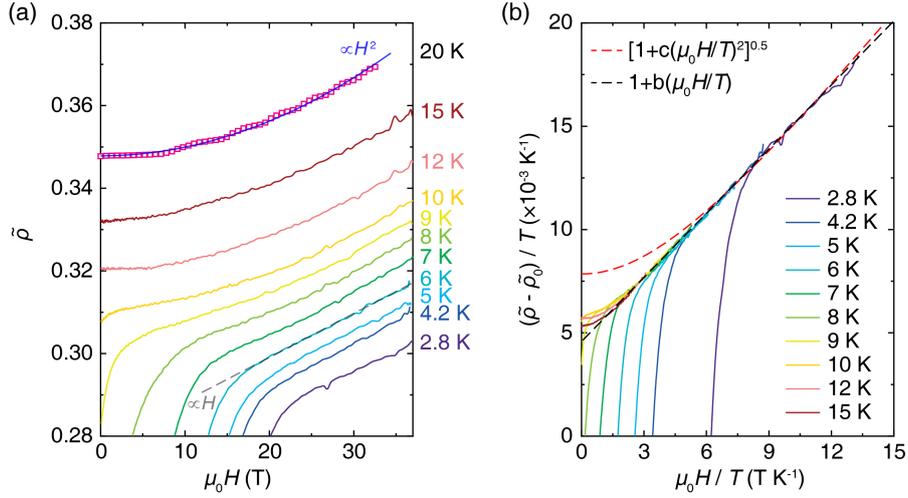

FIG. 4. Scaling between magnetic field and temperature. (a) Resistivity as a function of magnetic field at different temperatures for a pristine FeSe film (sample #13) with $T_c \sim 10$ K. The resistivity is normalized by the value at 250 K: $\tilde{\rho} = \rho/\rho(250\ K)$. The data from 2.8 to 15 K are from the same sample #13 measured in a pulsed magnet. The data at 20 K (purple squares at the top) are from sample #11 measured in a steady magnetic. The dashed grey and blue solid lines are fittings using linear- and $H^2$-dependence, respectively. (b) The scaling plots of the MR curves in panel (a), where $\tilde{\rho}_0 = \rho(0,0)/\rho(250\ K, 0\ T)$ is the residual resistivity obtained by extrapolating the $H$-linear $\tilde{\rho}$ at finite temperatures to $H = 0$ and $T = 0$. The plots are fitted with the functions proportional to $[1 + b(\mu_0 H/T)]$(black dashed line) and $\sqrt{1 + c(\mu_0 H/T)^2}$ (red dashed line), respectively. $b$ and $c$ are numeric parameters.

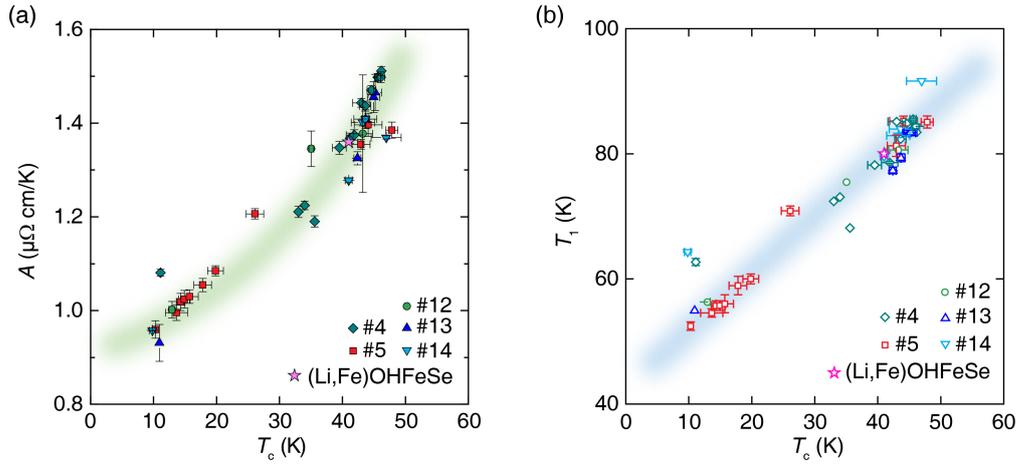

FIG. 5. The correlation between linear-in-$T$ resistivity and superconductivity. (a) The extracted linear resistivity coefficient $A$ as a function of $T_c$ for different samples. The $A$ is equal to unity for sample #5, and scaled to unity for other samples to remove variations due to geometric factor uncertainties. (b) The upper-bound temperature $T_1$ of the linear resistivity region as a function of $T_c$.